\documentclass[runningheads]{llncs}
\usepackage{comment}
\usepackage{graphicx}
\usepackage{acronym}
\usepackage{xspace}
\usepackage{multirow}
\usepackage[normalem]{ulem}
\useunder{\uline}{\ul}{}
\usepackage[table,xcdraw]{xcolor}
\usepackage[hyphens]{url} 
\usepackage[colorlinks=true,linkcolor=black,urlcolor=blue]{hyperref}
\newacro{PRNU}{Photo-Response Non Uniformity}
\newacro{PNU}{Pixel Non Uniformity}
\newacro{SCI}{Source Camera Identification}
\newacro{RN}{Residual Noise}
\newacro{RP}{Reference Pattern}
\newacro{SPN}{Sensor Pattern Noise}
\newacro{FPN}{Fixed Pattern Noise}
\newacro{CCN}{Circular Cross-Correlation Norm}
\newacro{CNCPO}{Centro Nazionale per il Contrasto alla Pedopornografia Online}
\newacro{LEA}{Law Enforcement Agency}
\newacroplural{LEAs}{Law Enforcement Agencies}

\newcommand{\etal}[1]{\emph{#1 et al.}\xspace}

\begin{document}
\title{On the Reliability of the PNU for Source Camera Identification Tasks}
%
%
\author{Andrea Bruno\inst{1}\orcidID{0000-0002-1821-7908} \and
Giuseppe Cattaneo\inst{1}\orcidID{0000-0002-6983-4818} \and
Paola Capasso\inst{1}
}
 \authorrunning{A. Bruno et al.}
%
\institute{Università degli Studi di Salerno, Dipartimento di Informatica, ITA 
\email{andbruno,cattaneo@unisa.it}\\
\url{ifaselab.di.unisa.it/}}
\maketitle              
%


\begin{abstract}
The \ac{PNU}\cite{1634362} is an essential and reliable tool to perform \ac{SCI} and, during the years, became a standard de-facto for this task in the forensic field. In this paper, we show that, although strategies exist that aim to cancel, modify, replace the \ac{PNU} traces in a digital camera image, it is still possible, through our experimental method, to find residual traces of the noise produced by the sensor used to shoot the photo. Furthermore, we show that is possible to inject the \ac{PNU} of a different camera in a target image and trace it back to the source camera, but only under the condition that the new camera is of the same model of the original one used to take the target image. Both cameras must fall within our availability. 

For completeness, we carried out $2$ experiments and, rather than using the popular public reference dataset, CASIA TIDE, we preferred to introduce a dataset that does not present any kind of statistical artifacts.
A preliminary experiment on a small dataset of smartphones showed that the injection of \ac{PNU} from a different device makes it impossible to identify the source camera correctly.

For a second experiment, we built a large dataset of images taken with the same model DSLR. We extracted a denoised version of each image, injected each one with the \ac{RN} of all the cameras in the dataset and compared all with a \ac{RP} from each camera. The results of the experiments, clearly, show that either in the denoised images and the injected ones is possible to find residual traces of the original camera \ac{PNU}.

The combined results of the experiments show that, even in theory is possible to remove or replace the \ac{PNU} from an image, this process can be, easily, detected and is possible, under some hard conditions, confirming the robustness of the \ac{PNU} under this type of attacks.

\keywords{Pixel Non Uniformity \and Source Camera Identification \and Anti-forensics.}
\end{abstract}
%
%

%


\section{Introduction}
\label{sec:intro}

The explosive growth of digital cameras and of their pervasive applications has led to an exponential increase of cases where the analysis of digital images plays an important role in crime investigations. Consequently, a new branch of computer forensic science has been introduced under the name of {\em Digital Image Forensics}.

A frequent problem that arises when conducting this type of investigation is about the identification of the camera that has been used to take one or more digital images under scrutiny. We define this problem as the {\em Source Camera Identification} problem (SCI, for short). A popular approach for the solution of this problem is to resort to the analysis of the {\em Sensor Pattern Noise}. This is a characteristic noise that is left in a systematic way by digital sensors when acquiring new images. As widely discussed by the several publications in this field (see, e.g.,  \cite{1634362}), it is possible to use this noise to determine a sort of unique fingerprint corresponding to the originating digital sensor.

 
The reliability of this approach has been assessed in several experimental studies. As a relevant result, it has been shown in \cite{Redi2011} that, assuming the originating camera $\mathcal{C}$ of a digital image $\mathcal{I}$ under scrutiny is available to an investigator, it is possible to identify $\mathcal{C}$ as the source camera for $\mathcal{I}$ with a very high level of confidence.

In the recent years, several {\em counter-forensics} techniques have been proposed for deceiving source camera identification. Here, the goal may be to prevent the identification of $\mathcal{C}$ (e.g., \cite{8296536}) or to modify $\mathcal{I}$ so as to make it result as taken by a different camera (e.g., ). 

In this paper, we focus on one of the most interesting {\em counter-forensics} technique, by describing a simple methodology for implementing the spoofing of a digital image. Our experimental results show that the proposed implementation succeeds in modifying a digital image so to make it result as taken using a camera different than its originating one. More surprisingly, our results also show that the filtering procedure, that is commonly used to remove from a digital image the traces about its originating camera, is partially uneffective. Consequently, it is still possible to reveal the source camera used to take a spoofed picture by means of a simple modification of the original source camera identification algorithm, thus confirming the robustness of the PRNU noise as a mean to perform source camera identification.





\section{PRNU-based Source Camera Identification}
\label{sec:tech}
In this section, we briefly outline one of the most popular Source Camera Identification technique, originally introduced by \etal{Fridrich} in \cite{1634362}. 

\label{sec:PR}

Their technique starts from the observation that the pattern noise of a digital image includes two main components: the \ac{FPN} noise and the Photo-Response Non Uniformity (\ac{PRNU}) noise. 
\ac{FPN} is an additive noise caused by dark currents and, also, depending on exposure and temperature 

\ac{PRNU} noise is a multiplicative noise dependent and is composed of :

\begin{itemize}
\item low frequency components or defects (due to, e.g., light refraction on dust particles, optical surfaces, zoom settings). They are of low spatial frequency in nature; therefore, they cannot be considered a characteristic of the sensor; 


\item \ac{PNU} noise, defined as the different sensitivity of pixels to light caused by the inhomogenity of silicon wafers used to manufacture digital sensors as well as other imperfections arising during the sensor manufacturing process. \end{itemize}

The key observation is that is very unlikely that different sensors, even manufactured using the same wafer, would exhibit correlated \ac{PNU} patterns. So, the \ac{PNU} noise is an intrinsic characteristic of each sensor and, thus, it can be effectively used for sensor identification.

Let $I$ be a digital image under scrutiny, the \etal{Fridrich} source camera identification technique can be summarized as follows:

\paragraph{Step 1} A {\em residual noise} $x$ is extracted from $I$ 
\begin{equation}
             x = I - F(I)\label{eq:rn}
\end{equation}

where $F$ is a denoising function. This is an approximation of the noise existing in $I$, including the PRNU noise. We will denote this residual noise as $x^{(I)}$. 

\paragraph{Step 2} \ac{PRNU} pattern of a reference camera was estimated from a series of photos taken with this camera. We adopt the maximum likelihood approach \cite{4451084} to estimate the camera's \ac{PRNU} pattern:

\begin{equation}
             K = \frac{ \sum_{i=1}^{N}{x^{(i)} I^{(i)}}}{\sum_{i=1}^N{(I^{(i)})^2}} \label{eq:rp}
\end{equation}
where $N$ is the series of images used to extract \ac{PRNU} $K$.

\paragraph{Step 3}  A correlation statistic is used to measure the similarity between the $x$ and $K$. In this document, we use a correlation statistic called correlation on \ac{CCN} \cite{6020790}, which is defined as:
 
\begin{equation}
             c(x,y) = \frac{xy/L}{\sqrt{\frac{1}{L-|A|}\sum_{m\notin A}r^2_{xy}(m)}}
             =
             \frac{r_{xy}(0)}{\sqrt{\frac{1}{L-|A|}\sum_{m\notin A}r^2_{xy}(m)}} 
             \label{eq:cc}
\end{equation}
where $y = KI$, $A$ is a small neighbourhood around zero displacement, $\big |A\big|$ is the measurement of $A$, and the \ac{CCN} $r_{xy}(m)$ is defined as 
\begin{equation}
              r_{xy} (m) =  \frac {1}{L} \sum_{I=0}^{L-1} x_{l y_l\oplus_m} \label{eq:ccn}
\end{equation}
where the operation $\oplus$ is the module $N$ addition. The investigator identifies the camera source by comparing the \ac{CCN} value with a predefined threshold. The higher the \ac{CCN} value, the more likely the test image is taken by the reference camera.

\section{Deceiving PRNU-based Source Camera Identification}
\label{sec:soa}

\subsection{Source Camera Identification Counter-forensics Techniques}

Counter-forensics techniques aim at making the analysis of evidence difficult or impossible to be carried out or, at least, they try to negatively affect the existence and the quality of evidences. In this work, we focus on counter-forensics techniques targeting source camera identification methodologies. According to literature, these can be grouped in the following cases:

 
\begin{description}
    \item[Fingerprint-Removal]: in order to prevent the identification of the source camera $\mathcal{C}$ used to take a digital image $\mathcal{I}$, the fingerprint of $\mathcal{C}$ in $\mathcal{I}$ is removed or altered;
    
    \item[Fingerprint-Injection]: in order to make a digital image $\mathcal{I}$ result to have been taken using a  camera $\mathcal{C^{'}}$ different than the original one, the fingerprint of $\mathcal{C^{'}}$ is properly injected in $\mathcal{I}$;
        
    \item[Fingerprint-Substitution]: a mix of the two previous attacks where, the fingerprint  of the source camera $\mathcal{C}$ used to take a digital image $\mathcal{I}$ is removed from $\mathcal{I}$ and replace with that of a different camera $\mathcal{C^{'}}$. This is attack is particularly useful to produce false evidences.

\end{description}





\subsubsection{Fingerprint-Remove Attacks}


    \paragraph{Basic Fingerprint Removal}: it estimates the model and the entity of the intrinsic strength of the PRNU of a picture and, then, subtracts it from that picture \cite{7351088}. 
	\paragraph{Adaptive Denoising \ac{PRNU} (ADP)}: It repeatedly applies a denoising filter to an image until it has sufficiently suppressed  its \ac{PRNU} noise so to prevent its source mapping. The goal is to get an image which would correlate very poorly with its own PRNU noise pattern \cite{KARAKUCUK201566}


    \paragraph{Seam-Carving}: 
    It is a content-sensitive image scaling technique used to disturb the reference noise pattern of an image.
    In this technique, a {\em seam}, which is a horizontal or vertical linked path of low-energy pixels,
    is removed or inserted into the image in a forced manner (at least one carved seam in a n x n block).
    When a seam is removed, the remaining pixels are shifted to fill the gap.
    The main idea behind image resizing based on seam engraving is to remove unnecessary parts of the image while keeping important content intact \cite{8006244}. 
    

\subsubsection{Fingerprint-Injection Attacks}

    \paragraph{Fingerprint-Copy Attack}: 
    In this technique, it is first estimated the fingerprint of a camera $C_{1}$ using a collection of images (e.g., a set of stolen images). Then, it is superimposed into a target image taken by a different camera $C_{2}$  to disguise the resulting image as one taken by $C_{1}$ \cite{6123122,5667057}. 
    

\subsubsection{Fingerprint-Substitution Attacks}
    \paragraph{Anonymization}: the technique misleads the identification of a camera $C_{1}$ based on the noise model of the PRNU sensor, using a median filter to suppress the PRNU noise existing in an image $I$ taken with $C_{1}$. 
    
    The PRNU of $I$ is removed and the variance of the PRNU of another camera $C_{2}$ is introduced to make the identity anonymous. The identity of $I$ is forged in such a way that it now appears that the image was produced by $C_{2}$ rather than $C_{1}$  or the counterfeit image has small traces of the original capture device \cite{raj2019counter,villalba2017prnu}

\subsection{Contrasting Counter-Forensic Techniques}

Source camera identification techniques are generally robust against preliminary image modification attacks and compression. Yet, they may fail when facing attacks like the ones reported in Section \ref{sec:soa}. For this reason, several methods have been developed for 
contrasting the effects of these attacks. 

Currently, the most relevant ones are the following:
\begin{description}
    \item[Demosaicing-based camera model identification:] this method detects if the fingerprint of the originating source camera embedded in a target image has been falsified. The algorithm works by characterizing the different local pixel relationships regardless of the content introduced by both authentic demosaicing algorithms and by anti-forensic attacks.
    An anti-forensic attacker, in fact, can falsify these traces by maliciously using existing forensic techniques to estimate one camera's demosaicing filter, then use these estimates to re-demosaic an image captured by another camera \cite{8296534}.

    \item[PRNU-based Forged Regions Detection:] this method detect and operate small forged regions and automatically. The \ac{SPN} is extracted from the target image and, then, it is correlated with the reference \ac{SPN} of a target camera. The two noises are divided into non-overlapping blocks before evaluating their correlations, so as to return a correlation map. Then, a set of operators is applied on the resulting  map to highlight the forged regions and remove the noise peaks \cite{10.1007/978-3-319-25903-1_42}.
    
    \item[The Triangle Test:]  this method has been proposed by \etal{Fridrich}  to reveal fingerprint-copy attacks\cite{5667057}. It can be summarized as follows. 
        Let Alice and Eve be the victim and the attacker respectively. First, Alice publicly shares a collection of images. Then, Eve grabs a copy of the images shared by Alice and uses them to extract the fingerprint of the camera used to take them. Given that the digital fingerprint of the camera estimated by Eva must be of high quality and estimated by at least 300 images, its estimation error contains the residuals of the entire residual noise of all the images used. 
        Using the correlation method, Alice can identify the images used by Eve for deriving the target fingerprint used for the falsification and, so, she can prove her innocence. Alice will be successful even when, being unable to arrange any of the images, she will be able to analyze at least two counterfeit images from Eve.
\end{description}

\section{Experimental analysis} 
\label{sec:exp}


Our thesis has been subject to a thorough experimental analysis.The goal has been to assess if: 1) there are traces of the originating camera for an image being spoofed 2) these traces are relevant enough to allow to identify the originating camera. 

To obtain more realistic results, we used different tools to carry on the attack and to analyze its effect. This is a realistic scenario, because is highly probable that the tools used by the attacker to extract the \ac{PRNU} from a collection of images, compute the corresponding \ac{RP} and alter the original images are different from the ones used for later analysis. In the same time, the \newacro{LEA}{} will probably uses some standard library for the same purpose. For both roles, we resorted to tools implementing the algorithm by \etal{Fridrich} in \cite{1634362}, already available in literature and widely used, as follows.
\begin{description}
    \item[Attacker:] here we used the matlab library developed at the Computer Science Department of Università degli Studi di Salerno and publicly available on GitLab \url{https://gitlab.com/dif_unisa/pnu_matlab}. It has been used in several publications, like \cite{10.1007/978-3-319-25903-1_42,10.1007/978-3-319-48680-2_64,10.1007/978-3-642-55032-4_66,10.1145/3264746.3264770,10.1007/978-3-319-70742-6_32,10.1007/978-3-030-30143-9_26,doi:10.1504/IJES.2020.107641}
    \item[Defender:] here we used the matlab library  developed by the DDE Laboratory at Binghamton University. It has been used in several publications, like \etal{Fridrich}\cite{10.1117/12.805701,10.1117/12.766732,10.1117/12.839055,5667057,10.1117/12.838378,10.1117/12.872198,10.1117/12.909659,Goljan:2016:2470-1173:1}
\end{description}

\subsection{Dataset}
\label{sub:dataset}
The ideal dataset to be used for our experiments should be made of images taken in such a way to not simplify neither the identification of the original cameras neither the spoofing process. For this reason, we opted for not using the standard CASIA TIDE (Tampered Image Detection Evaluation) dataset, as it has proven to include images containing statistical artifacts able to influence the identification process (see \cite{cattaneo2014possible}).

\begin{figure}[htpb]
\includegraphics[width=.9\textwidth]{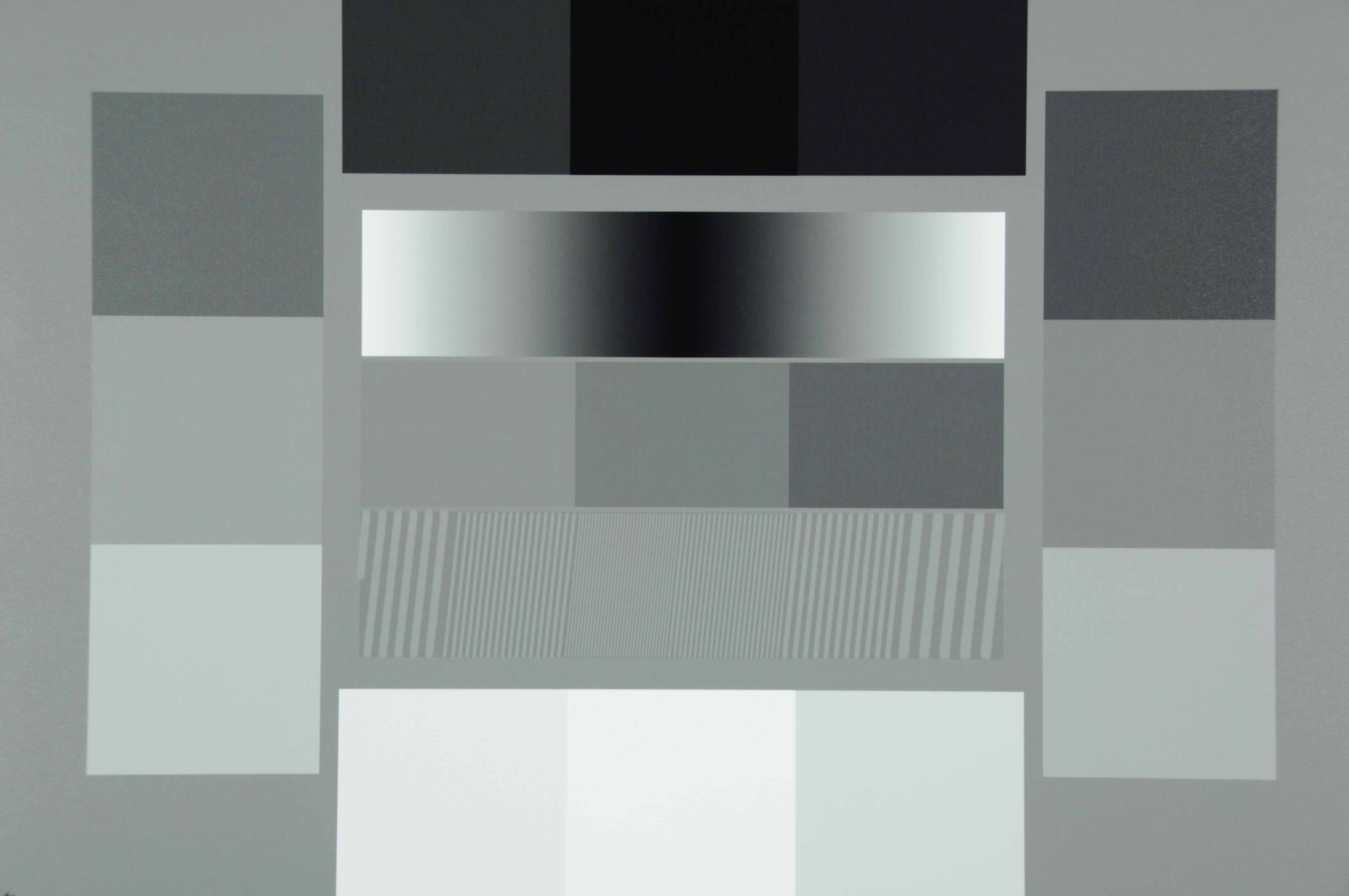}
\centering
\label{fig:iso15739example}
\caption{Example of enrollment images taken with Nikon D90 and ISO 15739:2017 noise enhancement sheet}
\end{figure}

Instead, we created two reference dataset:
\begin{itemize}
    \item {\bf small} dataset: it includes $1,000$ images coming from $4$ different smartphone cameras. The structure of this dataset is reported in table \ref{tab:dataset1};
    \item {\bf large} dataset: it includes $5,140$ images coming from $20$ different cameras. Each of these images has been taken taken using camera tripod, controlled light and, as subject, a sheet conforming to ISO 15739:2017 (see Fig \ref{fig:iso15739example}) to maximize the \ac{PRNU} in the images. This dataset has been assembled in cooperation with the Italian Postal Police Department for fight against pedopornography.The structure of this dataset is reported in table \ref{tab:dataset2}.
\end{itemize}

\begin{table}[htpb]
\centering
\resizebox{\textwidth}{!}{%
\begin{tabular}{|l|l|l|l|l|}
\hline
           & \textbf{Brand} & \textbf{Model} & \textbf{Resolution}   & \textbf{\#images} \\ \hline
\textbf{1} & Samsung        & Galaxy S9 Plus & $4032 \times 3024$ px & $250$             \\ \hline
\textbf{2} & Honor          & Honor 8        & $3968 \times 3024$ px & $250$             \\ \hline
\textbf{3} & Apple          & iPhone Xs      & $4032 \times 3024$ px & $250$             \\ \hline
\textbf{4} & Apple          & iPhone SE      & $4290 \times 2800$ px & $250$             \\ \hline
\end{tabular}%
}
\caption{Structure of our {\bf small} dataset}
\label{tab:dataset1}
\end{table}

\begin{table}[htpb]
\centering
\resizebox{\textwidth}{!}{%
\begin{tabular}{|l|l|l|l|c|c|c|c|}
\hline
\textbf{Identifier} & \textbf{Brand} & \textbf{Model} & \textbf{Resolution} & \textbf{Inject RP} & \textbf{Comparison RP} & \textbf{Spoofable} & \textbf{Total} \\ \hline
\textbf{IC\_190} & Nikon & D90  & $4288 \times 2848$ & $100$ & $100$ & $57$ & $257$ \\ \hline
\textbf{IC\_191} & Nikon & D91  & $4288 \times 2848$ & $100$ & $100$ & $57$ & $257$ \\ \hline
\textbf{IC\_192} & Nikon & D92  & $4288 \times 2848$ & $100$ & $100$ & $57$ & $257$ \\ \hline
\textbf{IC\_193} & Nikon & D93  & $4288 \times 2848$ & $100$ & $100$ & $57$ & $257$ \\ \hline
\textbf{IC\_194} & Nikon & D94  & $4288 \times 2848$ & $100$ & $100$ & $57$ & $257$ \\ \hline
\textbf{IC\_195} & Nikon & D95  & $4288 \times 2848$ & $100$ & $100$ & $57$ & $257$ \\ \hline
\textbf{IC\_196} & Nikon & D96  & $4288 \times 2848$ & $100$ & $100$ & $57$ & $257$ \\ \hline
\textbf{IC\_197} & Nikon & D97  & $4288 \times 2848$ & $100$ & $100$ & $57$ & $257$ \\ \hline
\textbf{IC\_198} & Nikon & D98  & $4288 \times 2848$ & $100$ & $100$ & $57$ & $257$ \\ \hline
\textbf{IC\_199} & Nikon & D99  & $4288 \times 2848$ & $100$ & $100$ & $57$ & $257$ \\ \hline
\textbf{IC\_200} & Nikon & D100 & $4288 \times 2848$ & $100$ & $100$ & $57$ & $257$ \\ \hline
\textbf{IC\_201} & Nikon & D101 & $4288 \times 2848$ & $100$ & $100$ & $57$ & $257$ \\ \hline
\textbf{IC\_202} & Nikon & D102 & $4288 \times 2848$ & $100$ & $100$ & $57$ & $257$ \\ \hline
\textbf{IC\_203} & Nikon & D103 & $4288 \times 2848$ & $100$ & $100$ & $57$ & $257$ \\ \hline
\textbf{IC\_204} & Nikon & D104 & $4288 \times 2848$ & $100$ & $100$ & $57$ & $257$ \\ \hline
\textbf{IC\_205} & Nikon & D105 & $4288 \times 2848$ & $100$ & $100$ & $57$ & $257$ \\ \hline
\textbf{IC\_206} & Nikon & D106 & $4288 \times 2848$ & $100$ & $100$ & $57$ & $257$ \\ \hline
\textbf{IC\_207} & Nikon & D107 & $4288 \times 2848$ & $100$ & $100$ & $57$ & $257$ \\ \hline
\textbf{IC\_208} & Nikon & D108 & $4288 \times 2848$ & $100$ & $100$ & $57$ & $257$ \\ \hline
\textbf{IC\_209} & Nikon & D109 & $4288 \times 2848$ & $100$ & $100$ & $57$ & $257$ \\ \hline
\end{tabular}%
}
\caption{Structure of our {\bf large} dataset }
\label{tab:dataset2}
\end{table}


\subsection{Description of the experiment}

Let $\mathcal{C}$ and $\mathcal{C^{'}}$ be two different cameras and let $\mathcal{I}$ be an image taken using $\mathcal{C}$. Here the attacker is interested in spoofing $\mathcal{I}$ so to make it result as taken using $\mathcal{C^{'}}$.
We assume that the attacker does not own camera $\mathcal{C^{'}}$ but, instead, he has a collection of images taken with this camera.

\label{sub:experimentdescription}
We can define three steps in the experiment:
\begin{enumerate}
    \item Setup
    \item Spoofing
    \item Comparing
\end{enumerate}

The first and last step contains action that are commons to the attacker and the investigator, like the generation of the \acp{RP}, and must not rely on a particular implementation of the \ac{PRNU} extraction libraries.

In the first steps, we generate a couple of \acp{RP}, one used by the attacker for the spoofing procedure, call it $RP^s_D$ where $D$ is the device of the $RP$, and one used by the investigator for the comparison procedure, call it $RP^c_D$ where $D$ is the device of the $RP$, for each of the camera. For the first experiment, we generate $8$ \ac{RP} processing $800$ images, for the second experiment $40$ \acp{RP} processing $4000$ images.

In the second step, we filter each of the $50$ target images of each camera with a Daubechis 8 wavelet filter to obtain a noiseless version $F_I$ of the image $F$.

To each $F_I$, we sum the $RP^s_D$ for each of the device $D$ in the dataset used for the experiment and the result is normalized in the range $[0,255]$ obtaining the image $I^D_{F_I}$

\begin{equation}
    I^D_{F_I} = F_I \oplus RP^s_D \quad \forall~D~\in~\textrm{Dataset}
\end{equation}

At the and of this step, we have $200$ filtered and $800$ spoofed images for the first dataset and $1140$ filtered and $22800$ spoofed images for the second dataset.

In the last step, we built a correlation matrix by correlating each one spoofed images with all the comparison \acp{RP} generated in step one.

\subsection{Experiment 1: Image Spoofing with crop and resize}
\label{sub:preliminaryexperiment}

In this first experiment, we are assuming that the image $\mathcal{I}$ being spoofed by the attacker has a different resolution than the one of the sensor equipped by $\mathcal{C^{'}}$. This can be assumed to be a sort of baseline as it is unlikely, in the average case, that these resolutions are identical.


Due the different size of the sensors of the different devices, the \acp{RP} must be cropped or resized in order to fit into the target images.

The results of this experiment are reported in Table  \ref{tab:spoof-experiment-results}. We consider the spoofing to be successful if the value of the correlation between the spoofing device and the spoofed images is the highest over the correlation between the images and the comparison \acp{RP}. Indeed, the spoofing activity is mostly unsuccessful, as in just the $5\%$ of cases the spoofed image is erroneously linked to a camera different than the originating one. It is very likely that this failure is mostly due to the crop and resize operations applied to \acp{RP} before injecting it in the images being spoofed. These operations are known to affect the results of correlation between \acp{RP} and \acp{RN}, because of the misalignments introduced in their corresponding \ac{PNU}s.



\begin{table}[htpb]
\centering
\resizebox{\textwidth}{!}{%
\begin{tabular}{|l|c|c|c|}
\hline
\multirow{2}{*}{} & \multirow{2}{*}{{\ul \textbf{Small Dataset}}} & \multicolumn{2}{c|}{{\ul \textbf{Large Dataset}}} \\ \cline{3-4} 
                                          &                   & \textbf{Full} & \textbf{Spoofed} \\ \hline
\textbf{\# of devices}                     & $4$               & \multicolumn{2}{c|}{$20$}                                                \\ \hline
\textbf{\# of training images per device} & $50$              & \multicolumn{2}{c|}{$57$}                                                \\ \hline
\textbf{\# of spoofed images}                & $800$             & $23940$                          & $22800$                               \\ \hline
\textbf{Percentage of successful spoofing}              & $229$ / $28.63\%$ & $21048$ / $87.92\%$              & $21048$ / $92.32\%$                   \\ \hline
\textbf{Percentage of failed spoofing}            & $571$ / $71.38\%$ & $2932$ / $12.08\%$               & $1752$ / $7.68\%$                     \\ \hline
\end{tabular}%
}
\caption{Result of the spoofing experiment on both datasets. For the large dataset, {\bf Full} includes all the images, while {\bf Spoofed} includes only the images spoofed with other cameras }
\label{tab:spoof-experiment-results}
\end{table}


\subsection{Experiment 2: Image Spoofing without crop and resize}
\label{sub:mainexperiment}

We report the results for this second experiment in Table \ref{tab:spoof-experiment-results}. 

The success rate for this experiment is over $99\%$, a value that is inline with the success rate in standard \ac{SCI} over the same dataset (data not shown but available upon request). Thus, the spoofing activity seems to have been successful. However, if we look closer at these results, this does not seem to be true anymore. In Table 
\ref{tab:correlationheatmap} we report the mean of the correlations evaluated during this experiment, aggregated according to the originating camera and the camera being spoofed. On a side, as expected, the highest correlation is reached when comparing the PRNU noise of a spoofed image with the reference pattern of its corresponding spoofing camera. On the other side, we notice that the second highest correlation value observed in almost all cases is the one between the PRNU noise of a spoofed image and the reference pattern of its originating camera. Two are the main consequences of these results. The first is that the filtering operation applied to an image is not able to remove all traces of the originating camera. The second is that, despite the spoofing, it is still possible  to identify the camera used to take a picture with a high confidence level.



\begin{table}[htpb]
\centering
\resizebox{\textwidth}{!}{%
\begin{tabular}{cllllllllllllllllllll}
\textbf{} &
  \multicolumn{1}{c}{\textbf{IC\_190}} &
  \multicolumn{1}{c}{\textbf{IC\_191}} &
  \multicolumn{1}{c}{\textbf{IC\_192}} &
  \multicolumn{1}{c}{\textbf{IC\_193}} &
  \multicolumn{1}{c}{\textbf{IC\_194}} &
  \multicolumn{1}{c}{\textbf{IC\_195}} &
  \multicolumn{1}{c}{\textbf{IC\_196}} &
  \multicolumn{1}{c}{\textbf{IC\_197}} &
  \multicolumn{1}{c}{\textbf{IC\_198}} &
  \multicolumn{1}{c}{\textbf{IC\_199}} &
  \multicolumn{1}{c}{\textbf{IC\_200}} &
  \multicolumn{1}{c}{\textbf{IC\_201}} &
  \multicolumn{1}{c}{\textbf{IC\_202}} &
  \multicolumn{1}{c}{\textbf{IC\_203}} &
  \multicolumn{1}{c}{\textbf{IC\_204}} &
  \multicolumn{1}{c}{\textbf{IC\_205}} &
  \multicolumn{1}{c}{\textbf{IC\_206}} &
  \multicolumn{1}{c}{\textbf{IC\_207}} &
  \multicolumn{1}{c}{\textbf{IC\_208}} &
  \multicolumn{1}{c}{\textbf{IC\_209}} \\
\textbf{FILTR} &
  \cellcolor[HTML]{FDC97D}0,002125 &
  \cellcolor[HTML]{FEE282}0,003977 &
  \cellcolor[HTML]{FEE582}0,004208 &
  \cellcolor[HTML]{FEE182}0,003904 &
  \cellcolor[HTML]{FEDA80}0,003397 &
  \cellcolor[HTML]{FDC67C}0,001932 &
  \cellcolor[HTML]{FEE382}0,00407 &
  \cellcolor[HTML]{FEE683}0,004304 &
  \cellcolor[HTML]{FDD57F}0,003039 &
  \cellcolor[HTML]{FEDF81}0,003809 &
  \cellcolor[HTML]{FEE082}0,003861 &
  \cellcolor[HTML]{FDCA7D}0,002255 &
  \cellcolor[HTML]{FDD780}0,003175 &
  \cellcolor[HTML]{FED880}0,003266 &
  \cellcolor[HTML]{FEE683}0,004316 &
  \cellcolor[HTML]{FEE582}0,004224 &
  \cellcolor[HTML]{FDD47F}0,002975 &
  \cellcolor[HTML]{FEE482}0,00415 &
  \cellcolor[HTML]{FEE482}0,004161 &
  \cellcolor[HTML]{FDD67F}0,003081 \\
\textbf{IC\_190} &
  \cellcolor[HTML]{7CC57D}0,234142 &
  \cellcolor[HTML]{FEE683}0,004322 &
  \cellcolor[HTML]{FFEB84}0,005836 &
  \cellcolor[HTML]{FEE482}0,004118 &
  \cellcolor[HTML]{F98971}-0,00257 &
  \cellcolor[HTML]{FBA175}-0,00078 &
  \cellcolor[HTML]{FFEB84}0,006135 &
  \cellcolor[HTML]{FEE382}0,00409 &
  \cellcolor[HTML]{FBAB77}-2E-05 &
  \cellcolor[HTML]{FDD37F}0,002863 &
  \cellcolor[HTML]{FAA075}-0,00089 &
  \cellcolor[HTML]{FEDB81}0,003506 &
  \cellcolor[HTML]{FDCB7D}0,002321 &
  \cellcolor[HTML]{FEE082}0,003868 &
  \cellcolor[HTML]{FFEB84}0,005823 &
  \cellcolor[HTML]{FCBE7B}0,001352 &
  \cellcolor[HTML]{FFEB84}0,004723 &
  \cellcolor[HTML]{FFEB84}0,004864 &
  \cellcolor[HTML]{FEEB84}0,007153 &
  \cellcolor[HTML]{FED880}0,00329 \\
\textbf{IC\_191} &
  \cellcolor[HTML]{FDCF7E}0,002597 &
  \cellcolor[HTML]{87C97E}0,214915 &
  \cellcolor[HTML]{FEDD81}0,003638 &
  \cellcolor[HTML]{FEE983}0,00453 &
  \cellcolor[HTML]{FEE683}0,004276 &
  \cellcolor[HTML]{FFEB84}0,005304 &
  \cellcolor[HTML]{FFEB84}0,005545 &
  \cellcolor[HTML]{FFEB84}0,004746 &
  \cellcolor[HTML]{FEE082}0,003881 &
  \cellcolor[HTML]{FEE683}0,004269 &
  \cellcolor[HTML]{FFEB84}0,005426 &
  \cellcolor[HTML]{FFEB84}0,005243 &
  \cellcolor[HTML]{FFEB84}0,005524 &
  \cellcolor[HTML]{FEE081}0,003856 &
  \cellcolor[HTML]{FEEB84}0,006468 &
  \cellcolor[HTML]{FFEB84}0,006092 &
  \cellcolor[HTML]{FEE282}0,004011 &
  \cellcolor[HTML]{FEE983}0,004506 &
  \cellcolor[HTML]{FFEB84}0,005637 &
  \cellcolor[HTML]{FFEB84}0,005062 \\
\textbf{IC\_192} &
  \cellcolor[HTML]{FEDB80}0,003468 &
  \cellcolor[HTML]{FED880}0,003271 &
  \cellcolor[HTML]{85C87D}0,217339 &
  \cellcolor[HTML]{FCEB84}0,009874 &
  \cellcolor[HTML]{FDD47F}0,002983 &
  \cellcolor[HTML]{FDCC7E}0,002404 &
  \cellcolor[HTML]{FEDE81}0,003742 &
  \cellcolor[HTML]{FDD780}0,003153 &
  \cellcolor[HTML]{FCBB7A}0,001099 &
  \cellcolor[HTML]{FBA877}-0,00027 &
  \cellcolor[HTML]{FEEA83}0,004584 &
  \cellcolor[HTML]{FA9974}-0,00137 &
  \cellcolor[HTML]{F98370}-0,00299 &
  \cellcolor[HTML]{FFEB84}0,006206 &
  \cellcolor[HTML]{FFEB84}0,005882 &
  \cellcolor[HTML]{FEEB84}0,006678 &
  \cellcolor[HTML]{FEEB84}0,006631 &
  \cellcolor[HTML]{FDEB84}0,009673 &
  \cellcolor[HTML]{FFEB84}0,005232 &
  \cellcolor[HTML]{FEEB84}0,007723 \\
\textbf{IC\_193} &
  \cellcolor[HTML]{FDD07E}0,002643 &
  \cellcolor[HTML]{FEE883}0,004454 &
  \cellcolor[HTML]{FCEB84}0,010461 &
  \cellcolor[HTML]{82C77D}0,223698 &
  \cellcolor[HTML]{FA8E72}-0,00216 &
  \cellcolor[HTML]{FEDC81}0,003553 &
  \cellcolor[HTML]{FEEB84}0,00703 &
  \cellcolor[HTML]{FCB479}0,000626 &
  \cellcolor[HTML]{FBAF78}0,000274 &
  \cellcolor[HTML]{FDD57F}0,003018 &
  \cellcolor[HTML]{FDD37F}0,002927 &
  \cellcolor[HTML]{F97E6F}-0,00336 &
  \cellcolor[HTML]{FCB579}0,000651 &
  \cellcolor[HTML]{FFEB84}0,005068 &
  \cellcolor[HTML]{FFEB84}0,004663 &
  \cellcolor[HTML]{FEEB84}0,006571 &
  \cellcolor[HTML]{FDEB84}0,008525 &
  \cellcolor[HTML]{FEEB84}0,0075 &
  \cellcolor[HTML]{FEEB84}0,006764 &
  \cellcolor[HTML]{FFEB84}0,00616 \\
\textbf{IC\_194} &
  \cellcolor[HTML]{FBA376}-0,00066 &
  \cellcolor[HTML]{FFEB84}0,004808 &
  \cellcolor[HTML]{FFEB84}0,004688 &
  \cellcolor[HTML]{FCC37C}0,001709 &
  \cellcolor[HTML]{C6DB81}0,104204 &
  \cellcolor[HTML]{FDD37F}0,00287 &
  \cellcolor[HTML]{FEEA83}0,004584 &
  \cellcolor[HTML]{FEEB84}0,007509 &
  \cellcolor[HTML]{FEEB84}0,007999 &
  \cellcolor[HTML]{FEEB84}0,006539 &
  \cellcolor[HTML]{FDEB84}0,009471 &
  \cellcolor[HTML]{FEEB84}0,006507 &
  \cellcolor[HTML]{FFEB84}0,006307 &
  \cellcolor[HTML]{FDD780}0,00322 &
  \cellcolor[HTML]{FFEB84}0,005531 &
  \cellcolor[HTML]{FEEB84}0,007233 &
  \cellcolor[HTML]{FBA777}-0,00035 &
  \cellcolor[HTML]{FFEB84}0,005251 &
  \cellcolor[HTML]{FFEB84}0,005373 &
  \cellcolor[HTML]{FEEB84}0,007867 \\
\textbf{IC\_195} &
  \cellcolor[HTML]{FBAF78}0,000239 &
  \cellcolor[HTML]{FEEB84}0,006652 &
  \cellcolor[HTML]{FFEB84}0,004766 &
  \cellcolor[HTML]{FFEB84}0,004744 &
  \cellcolor[HTML]{FEE582}0,004226 &
  \cellcolor[HTML]{A9D380}0,154363 &
  \cellcolor[HTML]{FFEB84}0,005453 &
  \cellcolor[HTML]{FEDA80}0,003385 &
  \cellcolor[HTML]{FEE081}0,003825 &
  \cellcolor[HTML]{FED880}0,003296 &
  \cellcolor[HTML]{FEDE81}0,003719 &
  \cellcolor[HTML]{FEDB81}0,003487 &
  \cellcolor[HTML]{FFEB84}0,005256 &
  \cellcolor[HTML]{FEDA80}0,003408 &
  \cellcolor[HTML]{FFEB84}0,005892 &
  \cellcolor[HTML]{FFEB84}0,005695 &
  \cellcolor[HTML]{FFEB84}0,004683 &
  \cellcolor[HTML]{FEE081}0,003834 &
  \cellcolor[HTML]{FFEB84}0,005139 &
  \cellcolor[HTML]{FFEB84}0,005205 \\
\textbf{IC\_196} &
  \cellcolor[HTML]{FEE182}0,003933 &
  \cellcolor[HTML]{FFEB84}0,005792 &
  \cellcolor[HTML]{FEE783}0,004338 &
  \cellcolor[HTML]{FFEB84}0,006201 &
  \cellcolor[HTML]{FEE582}0,004188 &
  \cellcolor[HTML]{FEDF81}0,003803 &
  \cellcolor[HTML]{8BCA7E}0,20776 &
  \cellcolor[HTML]{FFEB84}0,005277 &
  \cellcolor[HTML]{FEDE81}0,003715 &
  \cellcolor[HTML]{FFEB84}0,005875 &
  \cellcolor[HTML]{FEE983}0,0045 &
  \cellcolor[HTML]{FEE582}0,004187 &
  \cellcolor[HTML]{FDEB84}0,008896 &
  \cellcolor[HTML]{FDC77D}0,002022 &
  \cellcolor[HTML]{FA9E75}-0,00104 &
  \cellcolor[HTML]{FFEB84}0,0055 &
  \cellcolor[HTML]{FFEB84}0,004645 &
  \cellcolor[HTML]{FDD37F}0,002925 &
  \cellcolor[HTML]{F9EA84}0,016204 &
  \cellcolor[HTML]{FDD27F}0,002856 \\
\textbf{IC\_197} &
  \cellcolor[HTML]{FDCD7E}0,002425 &
  \cellcolor[HTML]{FEE382}0,004061 &
  \cellcolor[HTML]{FDCA7D}0,002207 &
  \cellcolor[HTML]{FA9D75}-0,00106 &
  \cellcolor[HTML]{FCEB84}0,010031 &
  \cellcolor[HTML]{FBB279}0,000428 &
  \cellcolor[HTML]{FEEA83}0,004606 &
  \cellcolor[HTML]{63BE7B}0,276179 &
  \cellcolor[HTML]{FEEB84}0,007894 &
  \cellcolor[HTML]{FFEB84}0,005678 &
  \cellcolor[HTML]{FBEA84}0,012029 &
  \cellcolor[HTML]{FDEB84}0,008495 &
  \cellcolor[HTML]{FEE983}0,004519 &
  \cellcolor[HTML]{FDD37F}0,002882 &
  \cellcolor[HTML]{FEEB84}0,006905 &
  \cellcolor[HTML]{FEEB84}0,007182 &
  \cellcolor[HTML]{F8766D}-0,00397 &
  \cellcolor[HTML]{FFEB84}0,005151 &
  \cellcolor[HTML]{FFEB84}0,005201 &
  \cellcolor[HTML]{FCB679}0,000748 \\
\textbf{IC\_198} &
  \cellcolor[HTML]{FA9A74}-0,00128 &
  \cellcolor[HTML]{FFEB84}0,004728 &
  \cellcolor[HTML]{FDCB7D}0,002311 &
  \cellcolor[HTML]{FCB379}0,000523 &
  \cellcolor[HTML]{FAEA84}0,01414 &
  \cellcolor[HTML]{FDD57F}0,003042 &
  \cellcolor[HTML]{FEE783}0,004395 &
  \cellcolor[HTML]{FCEB84}0,010544 &
  \cellcolor[HTML]{6BC17C}0,2626 &
  \cellcolor[HTML]{FFEB84}0,005492 &
  \cellcolor[HTML]{FDEB84}0,008129 &
  \cellcolor[HTML]{FEEB84}0,008084 &
  \cellcolor[HTML]{FFEB84}0,00507 &
  \cellcolor[HTML]{FDD27F}0,002811 &
  \cellcolor[HTML]{FFEB84}0,006096 &
  \cellcolor[HTML]{FEEB84}0,007409 &
  \cellcolor[HTML]{FA9C74}-0,00116 &
  \cellcolor[HTML]{FEE082}0,003861 &
  \cellcolor[HTML]{FFEB84}0,005054 &
  \cellcolor[HTML]{FCEB84}0,010088 \\
\textbf{IC\_199} &
  \cellcolor[HTML]{FCBC7B}0,001225 &
  \cellcolor[HTML]{FEE883}0,004417 &
  \cellcolor[HTML]{FBAE78}0,000172 &
  \cellcolor[HTML]{FDD27F}0,002819 &
  \cellcolor[HTML]{FDEB84}0,008138 &
  \cellcolor[HTML]{FDC67C}0,001916 &
  \cellcolor[HTML]{FFEB84}0,006167 &
  \cellcolor[HTML]{FEEB84}0,006431 &
  \cellcolor[HTML]{FFEB84}0,004924 &
  \cellcolor[HTML]{72C37C}0,250346 &
  \cellcolor[HTML]{FFEB84}0,005102 &
  \cellcolor[HTML]{FEEB84}0,006843 &
  \cellcolor[HTML]{FCEB84}0,009925 &
  \cellcolor[HTML]{FDCC7E}0,002402 &
  \cellcolor[HTML]{FFEB84}0,005518 &
  \cellcolor[HTML]{FFEB84}0,004689 &
  \cellcolor[HTML]{FEDE81}0,003717 &
  \cellcolor[HTML]{FCBE7B}0,001363 &
  \cellcolor[HTML]{FFEB84}0,005917 &
  \cellcolor[HTML]{FDCC7E}0,002383 \\
\textbf{IC\_200} &
  \cellcolor[HTML]{F98871}-0,00265 &
  \cellcolor[HTML]{FFEB84}0,005623 &
  \cellcolor[HTML]{FEE683}0,00431 &
  \cellcolor[HTML]{FDD57F}0,00307 &
  \cellcolor[HTML]{FAEA84}0,01388 &
  \cellcolor[HTML]{FCB77A}0,000865 &
  \cellcolor[HTML]{FEE182}0,003939 &
  \cellcolor[HTML]{FCEA84}0,011236 &
  \cellcolor[HTML]{FEEB84}0,006938 &
  \cellcolor[HTML]{FFEB84}0,00467 &
  \cellcolor[HTML]{7BC57D}0,235594 &
  \cellcolor[HTML]{FDCF7E}0,002614 &
  \cellcolor[HTML]{FEE182}0,003942 &
  \cellcolor[HTML]{FFEB84}0,005226 &
  \cellcolor[HTML]{FFEB84}0,005004 &
  \cellcolor[HTML]{FDEB84}0,008775 &
  \cellcolor[HTML]{FCB679}0,000789 &
  \cellcolor[HTML]{FFEB84}0,005909 &
  \cellcolor[HTML]{FFEB84}0,005237 &
  \cellcolor[HTML]{FEE582}0,004192 \\
\textbf{IC\_201} &
  \cellcolor[HTML]{FDD27F}0,002832 &
  \cellcolor[HTML]{FFEB84}0,006106 &
  \cellcolor[HTML]{FCBE7B}0,001332 &
  \cellcolor[HTML]{FBA977}-0,00017 &
  \cellcolor[HTML]{FDEB84}0,00895 &
  \cellcolor[HTML]{FDD880}0,003255 &
  \cellcolor[HTML]{FFEB84}0,005424 &
  \cellcolor[HTML]{FDEB84}0,008254 &
  \cellcolor[HTML]{FEEB84}0,006477 &
  \cellcolor[HTML]{FEEB84}0,006828 &
  \cellcolor[HTML]{FFEB84}0,00463 &
  \cellcolor[HTML]{BCD881}0,122455 &
  \cellcolor[HTML]{FDEB84}0,008385 &
  \cellcolor[HTML]{FFEB84}0,004796 &
  \cellcolor[HTML]{FFEB84}0,0062 &
  \cellcolor[HTML]{FDD07E}0,002674 &
  \cellcolor[HTML]{FCBD7B}0,001291 &
  \cellcolor[HTML]{FBA676}-0,00045 &
  \cellcolor[HTML]{FFEB84}0,004829 &
  \cellcolor[HTML]{FEDC81}0,003528 \\
\textbf{IC\_202} &
  \cellcolor[HTML]{FCBD7B}0,001306 &
  \cellcolor[HTML]{FFEB84}0,005659 &
  \cellcolor[HTML]{FA8F72}-0,00209 &
  \cellcolor[HTML]{FCC17B}0,001551 &
  \cellcolor[HTML]{FDEB84}0,009028 &
  \cellcolor[HTML]{FEE883}0,004438 &
  \cellcolor[HTML]{FCEB84}0,010643 &
  \cellcolor[HTML]{FFEB84}0,005823 &
  \cellcolor[HTML]{FEE983}0,004532 &
  \cellcolor[HTML]{FCEB84}0,010196 &
  \cellcolor[HTML]{FEE883}0,004418 &
  \cellcolor[HTML]{FCEB84}0,009975 &
  \cellcolor[HTML]{85C87D}0,218588 &
  \cellcolor[HTML]{F97F6F}-0,00333 &
  \cellcolor[HTML]{FED980}0,003364 &
  \cellcolor[HTML]{FFEB84}0,005828 &
  \cellcolor[HTML]{FDD17F}0,002741 &
  \cellcolor[HTML]{FCB579}0,000649 &
  \cellcolor[HTML]{FDEB84}0,008424 &
  \cellcolor[HTML]{FFEB84}0,005018 \\
\textbf{IC\_203} &
  \cellcolor[HTML]{FDD07E}0,002686 &
  \cellcolor[HTML]{FFEB84}0,004669 &
  \cellcolor[HTML]{FEEB84}0,00746 &
  \cellcolor[HTML]{FFEB84}0,004805 &
  \cellcolor[HTML]{FDD17F}0,002783 &
  \cellcolor[HTML]{FDCD7E}0,002422 &
  \cellcolor[HTML]{FCC07B}0,001482 &
  \cellcolor[HTML]{FEEA83}0,004599 &
  \cellcolor[HTML]{FDD17F}0,002717 &
  \cellcolor[HTML]{FDD57F}0,003021 &
  \cellcolor[HTML]{FFEB84}0,005859 &
  \cellcolor[HTML]{FFEB84}0,004769 &
  \cellcolor[HTML]{F97D6E}-0,00343 &
  \cellcolor[HTML]{93CC7E}0,193877 &
  \cellcolor[HTML]{FEEB84}0,006382 &
  \cellcolor[HTML]{FEE683}0,004329 &
  \cellcolor[HTML]{FFEB84}0,004955 &
  \cellcolor[HTML]{FEEB84}0,007442 &
  \cellcolor[HTML]{FDCD7E}0,002437 &
  \cellcolor[HTML]{FEE282}0,003977 \\
\textbf{IC\_204} &
  \cellcolor[HTML]{FEDC81}0,003567 &
  \cellcolor[HTML]{FFEB84}0,005666 &
  \cellcolor[HTML]{FFEB84}0,005715 &
  \cellcolor[HTML]{FEE482}0,004134 &
  \cellcolor[HTML]{FFEB84}0,004819 &
  \cellcolor[HTML]{FEDD81}0,003638 &
  \cellcolor[HTML]{FA9F75}-0,00091 &
  \cellcolor[HTML]{FEEB84}0,006635 &
  \cellcolor[HTML]{FEE983}0,004505 &
  \cellcolor[HTML]{FFEB84}0,004894 &
  \cellcolor[HTML]{FFEB84}0,005071 &
  \cellcolor[HTML]{FFEB84}0,005213 &
  \cellcolor[HTML]{FDD07E}0,002703 &
  \cellcolor[HTML]{FFEB84}0,005428 &
  \cellcolor[HTML]{A3D17F}0,16552 &
  \cellcolor[HTML]{FEE983}0,004516 &
  \cellcolor[HTML]{FDCA7D}0,002252 &
  \cellcolor[HTML]{FCEA84}0,010877 &
  \cellcolor[HTML]{FDD27F}0,002854 &
  \cellcolor[HTML]{FEE683}0,00427 \\
\textbf{IC\_205} &
  \cellcolor[HTML]{FBA376}-0,00062 &
  \cellcolor[HTML]{FFEB84}0,006084 &
  \cellcolor[HTML]{FEEB84}0,006626 &
  \cellcolor[HTML]{FEEB84}0,006379 &
  \cellcolor[HTML]{FEEB84}0,007962 &
  \cellcolor[HTML]{FEE883}0,004435 &
  \cellcolor[HTML]{FFEB84}0,00548 &
  \cellcolor[HTML]{FEEB84}0,006682 &
  \cellcolor[HTML]{FFEB84}0,005915 &
  \cellcolor[HTML]{FEE081}0,003823 &
  \cellcolor[HTML]{FDEB84}0,008354 &
  \cellcolor[HTML]{FBAB77}-8,2E-05 &
  \cellcolor[HTML]{FFEB84}0,005218 &
  \cellcolor[HTML]{FEDB80}0,003466 &
  \cellcolor[HTML]{FEE983}0,004511 &
  \cellcolor[HTML]{87C97E}0,214824 &
  \cellcolor[HTML]{FCB479}0,000607 &
  \cellcolor[HTML]{FFEB84}0,005724 &
  \cellcolor[HTML]{FEEA83}0,004595 &
  \cellcolor[HTML]{FDEB84}0,009337 \\
\textbf{IC\_206} &
  \cellcolor[HTML]{FEE282}0,004006 &
  \cellcolor[HTML]{FFEB84}0,004761 &
  \cellcolor[HTML]{FEEB84}0,007475 &
  \cellcolor[HTML]{FDEB84}0,009573 &
  \cellcolor[HTML]{F8696B}-0,00496 &
  \cellcolor[HTML]{FEE282}0,004014 &
  \cellcolor[HTML]{FFEB84}0,005697 &
  \cellcolor[HTML]{FA9774}-0,0015 &
  \cellcolor[HTML]{FA9974}-0,00138 &
  \cellcolor[HTML]{FEE081}0,003854 &
  \cellcolor[HTML]{FCBC7B}0,001226 &
  \cellcolor[HTML]{FBA977}-0,00022 &
  \cellcolor[HTML]{FDCD7E}0,002455 &
  \cellcolor[HTML]{FFEB84}0,005465 &
  \cellcolor[HTML]{FDD67F}0,003081 &
  \cellcolor[HTML]{FCBE7B}0,00137 &
  \cellcolor[HTML]{72C37C}0,250338 &
  \cellcolor[HTML]{FFEB84}0,005127 &
  \cellcolor[HTML]{FFEB84}0,006134 &
  \cellcolor[HTML]{FCC17C}0,001586 \\
\textbf{IC\_207} &
  \cellcolor[HTML]{FDD27F}0,002846 &
  \cellcolor[HTML]{FEE683}0,004285 &
  \cellcolor[HTML]{FDEB84}0,009573 &
  \cellcolor[HTML]{FEEB84}0,007383 &
  \cellcolor[HTML]{FFEB84}0,004882 &
  \cellcolor[HTML]{FCC57C}0,001839 &
  \cellcolor[HTML]{FCC37C}0,001743 &
  \cellcolor[HTML]{FFEB84}0,005762 &
  \cellcolor[HTML]{FDD07E}0,002644 &
  \cellcolor[HTML]{FCBB7A}0,001159 &
  \cellcolor[HTML]{FFEB84}0,005804 &
  \cellcolor[HTML]{F86B6B}-0,00474 &
  \cellcolor[HTML]{FBA576}-0,00053 &
  \cellcolor[HTML]{FFEB84}0,006249 &
  \cellcolor[HTML]{FBEA84}0,013118 &
  \cellcolor[HTML]{FFEB84}0,006332 &
  \cellcolor[HTML]{FEE182}0,003917 &
  \cellcolor[HTML]{79C57D}0,238089 &
  \cellcolor[HTML]{FDC87D}0,00211 &
  \cellcolor[HTML]{FFEB84}0,005685 \\
\textbf{IC\_208} &
  \cellcolor[HTML]{FFEB84}0,004825 &
  \cellcolor[HTML]{FFEB84}0,005138 &
  \cellcolor[HTML]{FFEB84}0,005029 &
  \cellcolor[HTML]{FEEB84}0,006384 &
  \cellcolor[HTML]{FFEB84}0,005202 &
  \cellcolor[HTML]{FDD57F}0,003077 &
  \cellcolor[HTML]{F8E984}0,017009 &
  \cellcolor[HTML]{FFEB84}0,005224 &
  \cellcolor[HTML]{FEE182}0,003928 &
  \cellcolor[HTML]{FFEB84}0,00519 &
  \cellcolor[HTML]{FEEA83}0,004618 &
  \cellcolor[HTML]{FDD57F}0,003039 &
  \cellcolor[HTML]{FEEB84}0,007132 &
  \cellcolor[HTML]{FDC87D}0,002066 &
  \cellcolor[HTML]{FDCA7D}0,002264 &
  \cellcolor[HTML]{FEEA83}0,004615 &
  \cellcolor[HTML]{FEEA83}0,004625 &
  \cellcolor[HTML]{FDCB7D}0,00232 &
  \cellcolor[HTML]{92CC7E}0,194705 &
  \cellcolor[HTML]{FFEB84}0,005239 \\
\textbf{IC\_209} &
  \cellcolor[HTML]{FDCB7D}0,002291 &
  \cellcolor[HTML]{FFEB84}0,00603 &
  \cellcolor[HTML]{FDEB84}0,009133 &
  \cellcolor[HTML]{FEEB84}0,006759 &
  \cellcolor[HTML]{FBEA84}0,012402 &
  \cellcolor[HTML]{FFEB84}0,004723 &
  \cellcolor[HTML]{FED980}0,003346 &
  \cellcolor[HTML]{FDD07E}0,002658 &
  \cellcolor[HTML]{FDEB84}0,009435 &
  \cellcolor[HTML]{FDD37F}0,002872 &
  \cellcolor[HTML]{FFEB84}0,00545 &
  \cellcolor[HTML]{FEDE81}0,003682 &
  \cellcolor[HTML]{FFEB84}0,004975 &
  \cellcolor[HTML]{FEE783}0,004341 &
  \cellcolor[HTML]{FFEB84}0,005523 &
  \cellcolor[HTML]{FCEB84}0,010146 &
  \cellcolor[HTML]{FCC27C}0,001616 &
  \cellcolor[HTML]{FEEB84}0,006822 &
  \cellcolor[HTML]{FEEB84}0,006589 &
  \cellcolor[HTML]{7FC67D}0,228341
\end{tabular}%
}
\caption{Correlation heatmap for the experiment run on the large dataset. On each column, it is reported each of the considered reference camera devices. On each row, it is reported each of the camera device used for spoofing. The generic $(i, j)$ entry of the heatmap reports the correlation value obtained when comparing the \acp{RP} of camera $i$ with the PRNU noise extracted from an image where it was previously injected the \acp{RP} of $j$. The horizontal row labeled as \textbf{FILTR} represents the correlation values obtained by comparing the collection of images obtained by each of the considered camera devices with their filtered counterparts.}
\label{tab:correlationheatmap}
\end{table}



\begin{table}[htpb]
\centering
\resizebox{\textwidth}{!}{%
\begin{tabular}{|c|l|l|l|l|l|l|l|l|l|l|l|l|l|l|l|l|l|l|l|l|}
\hline
\multicolumn{1}{|l|}{} &
  \multicolumn{1}{c|}{\textbf{IC\_190}} &
  \multicolumn{1}{c|}{\textbf{IC\_191}} &
  \multicolumn{1}{c|}{\textbf{IC\_192}} &
  \multicolumn{1}{c|}{\textbf{IC\_193}} &
  \multicolumn{1}{c|}{\textbf{IC\_194}} &
  \multicolumn{1}{c|}{\textbf{IC\_195}} &
  \multicolumn{1}{c|}{\textbf{IC\_196}} &
  \multicolumn{1}{c|}{\textbf{IC\_197}} &
  \multicolumn{1}{c|}{\textbf{IC\_198}} &
  \multicolumn{1}{c|}{\textbf{IC\_199}} &
  \multicolumn{1}{c|}{\textbf{IC\_200}} &
  \multicolumn{1}{c|}{\textbf{IC\_201}} &
  \multicolumn{1}{c|}{\textbf{IC\_202}} &
  \multicolumn{1}{c|}{\textbf{IC\_203}} &
  \multicolumn{1}{c|}{\textbf{IC\_204}} &
  \multicolumn{1}{c|}{\textbf{IC\_205}} &
  \multicolumn{1}{c|}{\textbf{IC\_206}} &
  \multicolumn{1}{c|}{\textbf{IC\_207}} &
  \multicolumn{1}{c|}{\textbf{IC\_208}} &
  \textbf{IC\_209} \\ \hline
\textbf{IC\_190} &
  \cellcolor[HTML]{C6EFCE}{\color[HTML]{006100} 0.253579} &
  -0.0004 &
  \cellcolor[HTML]{FFEB9C}{\color[HTML]{9C5700} 0.088895} &
  0.002456 &
  -0.00672 &
  -0.00447 &
  0.001298 &
  -0.00182 &
  -0.00632 &
  -0.00547 &
  \cellcolor[HTML]{FFC7CE}{\color[HTML]{9C0006} -0.00689} &
  -0.00356 &
  -0.00507 &
  0.002491 &
  0.002895 &
  -0.00182 &
  -0.00012 &
  0.003421 &
  0.002491 &
  0.001579 \\ \hline
\textbf{IC\_191} &
  -0.00161 &
  \cellcolor[HTML]{C6EFCE}{\color[HTML]{006100} 0.232298} &
  \cellcolor[HTML]{FFEB9C}{\color[HTML]{9C5700} 0.089123} &
  0.002947 &
  0.000895 &
  0.002368 &
  -0.00021 &
  -0.00153 &
  -0.00218 &
  \cellcolor[HTML]{FFC7CE}{\color[HTML]{9C0006} -0.00423} &
  1,75E-05 &
  -0.00146 &
  -0.00209 &
  0.002368 &
  0.003316 &
  0.003351 &
  -0.001 &
  0.00314 &
  0.000439 &
  0.003947 \\ \hline
\textbf{IC\_192} &
  -3,5E-05 &
  -0.00172 &
  \cellcolor[HTML]{C6EFCE}{\color[HTML]{006100} 0.314246} &
  \cellcolor[HTML]{FFEB9C}{\color[HTML]{9C5700} 0.009368} &
  -0.00161 &
  -0.00082 &
  -0.00202 &
  -0.00318 &
  -0.00493 &
  -0.00905 &
  -0.00091 &
  -0.00974 &
  \cellcolor[HTML]{FFC7CE}{\color[HTML]{9C0006} -0.01142} &
  0.005088 &
  0.002351 &
  0.003526 &
  0.002719 &
  0.008842 &
  -8,8E-05 &
  0.006439 \\ \hline
\textbf{IC\_193} &
  -0.00116 &
  -0.00046 &
  \cellcolor[HTML]{FFEB9C}{\color[HTML]{9C5700} 0.095491} &
  \cellcolor[HTML]{C6EFCE}{\color[HTML]{006100} 0.244088} &
  -0.00623 &
  0.000667 &
  0.001614 &
  -0.0057 &
  -0.00619 &
  -0.00565 &
  -0.00263 &
  \cellcolor[HTML]{FFC7CE}{\color[HTML]{9C0006} -0.0107} &
  -0.00714 &
  0.003368 &
  0.001491 &
  0.003807 &
  0.004684 &
  0.006825 &
  0.00186 &
  0.004947 \\ \hline
\textbf{IC\_194} &
  -0.00525 &
  -0.00047 &
  \cellcolor[HTML]{FFEB9C}{\color[HTML]{9C5700} 0.09786} &
  0.000316 &
  \cellcolor[HTML]{C6EFCE}{\color[HTML]{006100} 0.112789} &
  -0.0004 &
  -0.0016 &
  0.001316 &
  0.001965 &
  -0.00249 &
  0.004193 &
  -0.00147 &
  -0.00216 &
  0.001474 &
  0.001982 &
  0.004702 &
  \cellcolor[HTML]{FFC7CE}{\color[HTML]{9C0006} -0.00575} &
  0.004228 &
  -0.00053 &
  0.007035 \\ \hline
\textbf{IC\_195} &
  -0.00409 &
  0.002 &
  \cellcolor[HTML]{FFEB9C}{\color[HTML]{9C5700} 0.094088} &
  0.003386 &
  0.000439 &
  \cellcolor[HTML]{C6EFCE}{\color[HTML]{006100} 0.166228} &
  -0.00072 &
  -0.00304 &
  -0.00237 &
  \cellcolor[HTML]{FFC7CE}{\color[HTML]{9C0006} -0.00581} &
  -0.0016 &
  -0.00396 &
  -0.00265 &
  0.001895 &
  0.002474 &
  0.003 &
  -0.00025 &
  0.00214 &
  -0.00033 &
  0.003579 \\ \hline
\textbf{IC\_196} &
  0.000421 &
  0.000965 &
  \cellcolor[HTML]{FFEB9C}{\color[HTML]{9C5700} 0.091474} &
  0.004895 &
  0.001053 &
  0.000175 &
  \cellcolor[HTML]{C6EFCE}{\color[HTML]{006100} 0.22214} &
  -0.00067 &
  -0.0023 &
  -0.00254 &
  -0.00121 &
  -0.00286 &
  0.001526 &
  0.00014 &
  \cellcolor[HTML]{FFC7CE}{\color[HTML]{9C0006} -0.00516} &
  0.002684 &
  -0.00035 &
  0.001719 &
  0.012123 &
  0.001263 \\ \hline
\textbf{IC\_197} &
  -0.001 &
  -0.00033 &
  \cellcolor[HTML]{FFEB9C}{\color[HTML]{9C5700} 0.079123} &
  -0.00328 &
  0.007596 &
  -0.00298 &
  -0.00039 &
  \cellcolor[HTML]{C6EFCE}{\color[HTML]{006100} 0.297667} &
  0.002982 &
  -0.00163 &
  0.008088 &
  0.003088 &
  -0.0026 &
  0.001 &
  0.004228 &
  0.004667 &
  \cellcolor[HTML]{FFC7CE}{\color[HTML]{9C0006} -0.00891} &
  0.00393 &
  0.000596 &
  -0.00109 \\ \hline
\textbf{IC\_198} &
  -0.00518 &
  0.000123 &
  \cellcolor[HTML]{FFEB9C}{\color[HTML]{9C5700} 0.081544} &
  -0.00132 &
  0.011947 &
  -0.00018 &
  -0.00114 &
  0.005737 &
  \cellcolor[HTML]{C6EFCE}{\color[HTML]{006100} 0.281228} &
  -0.002 &
  0.003386 &
  0.001965 &
  -0.00188 &
  0.001246 &
  0.003351 &
  0.005175 &
  \cellcolor[HTML]{FFC7CE}{\color[HTML]{9C0006} -0.00642} &
  0.002544 &
  0.000526 &
  0.009263 \\ \hline
\textbf{IC\_199} &
  \cellcolor[HTML]{FFC7CE}{\color[HTML]{9C0006} -0.00256} &
  -0.00042 &
  \cellcolor[HTML]{FFEB9C}{\color[HTML]{9C5700} 0.081614} &
  0.001035 &
  0.005263 &
  -0.00137 &
  0.000667 &
  0.001018 &
  -0.00033 &
  \cellcolor[HTML]{C6EFCE}{\color[HTML]{006100} 0.267298} &
  1,75E-05 &
  0.000175 &
  0.00293 &
  0.000596 &
  0.002193 &
  0.001719 &
  -0.00084 &
  -0.00032 &
  0.000737 &
  0.00086 \\ \hline
\textbf{IC\_200} &
  \cellcolor[HTML]{FFC7CE}{\color[HTML]{9C0006} -0.007} &
  0.000982 &
  \cellcolor[HTML]{FFEB9C}{\color[HTML]{9C5700} 0.087228} &
  0.001246 &
  0.011439 &
  -0.00211 &
  -0.002 &
  0.006228 &
  0.001702 &
  -0.00335 &
  \cellcolor[HTML]{C6EFCE}{\color[HTML]{006100} 0.254684} &
  -0.00337 &
  -0.00335 &
  0.004421 &
  0.001509 &
  0.006509 &
  -0.00377 &
  0.004754 &
  -9,1E-20 &
  0.00307 \\ \hline
\textbf{IC\_201} &
  -0.0014 &
  0.001158 &
  \cellcolor[HTML]{FFEB9C}{\color[HTML]{9C5700} 0.092684} &
  -0.00179 &
  0.005246 &
  -0.00021 &
  -0.00049 &
  0.002421 &
  0.000439 &
  -0.00196 &
  -0.00077 &
  \cellcolor[HTML]{C6EFCE}{\color[HTML]{006100} 0.129123} &
  0.000491 &
  0.003368 &
  0.002719 &
  -0.00082 &
  \cellcolor[HTML]{FFC7CE}{\color[HTML]{9C0006} -0.00446} &
  -0.00247 &
  -0.00084 &
  0.001456 \\ \hline
\textbf{IC\_202} &
  -0.00233 &
  0.001053 &
  \cellcolor[HTML]{FFEB9C}{\color[HTML]{9C5700} 0.081719} &
  8,77E-05 &
  0.005719 &
  0.001351 &
  0.005737 &
  -3E-20 &
  -0.00121 &
  0.002491 &
  -0.00102 &
  0.004298 &
  \cellcolor[HTML]{C6EFCE}{\color[HTML]{006100} 0.233596} &
  \cellcolor[HTML]{FFC7CE}{\color[HTML]{9C0006} -0.00553} &
  -0.00014 &
  0.003018 &
  -0.00188 &
  -0.00096 &
  0.003789 &
  0.003772 \\ \hline
\textbf{IC\_203} &
  -0.00084 &
  -0.00051 &
  \cellcolor[HTML]{FFEB9C}{\color[HTML]{9C5700} 0.093351} &
  0.003456 &
  -0.00116 &
  -0.0007 &
  -0.0047 &
  -0.00149 &
  -0.0033 &
  -0.00561 &
  0.000684 &
  -0.00212 &
  \cellcolor[HTML]{FFC7CE}{\color[HTML]{9C0006} -0.01209} &
  \cellcolor[HTML]{C6EFCE}{\color[HTML]{006100} 0.210982} &
  0.003246 &
  0.001246 &
  0.00014 &
  0.007 &
  -0.00346 &
  0.002579 \\ \hline
\textbf{IC\_204} &
  -0.0004 &
  0.000614 &
  \cellcolor[HTML]{FFEB9C}{\color[HTML]{9C5700} 0.095263} &
  0.002614 &
  0.001386 &
  0.00014 &
  \cellcolor[HTML]{FFC7CE}{\color[HTML]{9C0006} -0.00793} &
  0.000579 &
  -0.00142 &
  -0.00407 &
  -0.00079 &
  -0.00219 &
  -0.00593 &
  0.004123 &
  \cellcolor[HTML]{C6EFCE}{\color[HTML]{006100} 0.17914} &
  0.001614 &
  -0.00286 &
  0.010456 &
  -0.00267 &
  0.00293 \\ \hline
\textbf{IC\_205} &
  -0.00468 &
  0.001526 &
  \cellcolor[HTML]{FFEB9C}{\color[HTML]{9C5700} 0.091386} &
  0.005193 &
  0.005 &
  0.001281 &
  -3,5E-05 &
  0.001018 &
  0.000246 &
  -0.00479 &
  0.003298 &
  \cellcolor[HTML]{FFC7CE}{\color[HTML]{9C0006} -0.00782} &
  -0.00256 &
  0.001965 &
  0.001193 &
  \cellcolor[HTML]{C6EFCE}{\color[HTML]{006100} 0.231474} &
  -0.00405 &
  0.004596 &
  -0.00049 &
  0.008333 \\ \hline
\textbf{IC\_206} &
  -0.00011 &
  5,26E-05 &
  \cellcolor[HTML]{FFEB9C}{\color[HTML]{9C5700} 0.088614} &
  0.008807 &
  \cellcolor[HTML]{FFC7CE}{\color[HTML]{9C0006} -0.00846} &
  0.000614 &
  0.000211 &
  -0.00821 &
  -0.00786 &
  -0.00411 &
  -0.00402 &
  -0.00782 &
  -0.00472 &
  0.003807 &
  1,22E-19 &
  -0.00163 &
  \cellcolor[HTML]{C6EFCE}{\color[HTML]{006100} 0.26907} &
  0.003912 &
  0.001368 &
  0.000193 \\ \hline
\textbf{IC\_207} &
  -0.00095 &
  -0.00086 &
  \cellcolor[HTML]{FFEB9C}{\color[HTML]{9C5700} 0.092947} &
  0.006456 &
  0.002088 &
  -0.00181 &
  -0.00418 &
  -8,8E-05 &
  -0.00333 &
  -0.00744 &
  0.000316 &
  \cellcolor[HTML]{FFC7CE}{\color[HTML]{9C0006} -0.01289} &
  -0.00867 &
  0.005 &
  0.010632 &
  0.003772 &
  -0.00088 &
  \cellcolor[HTML]{C6EFCE}{\color[HTML]{006100} 0.258561} &
  -0.00333 &
  0.004772 \\ \hline
\textbf{IC\_208} &
  0.001175 &
  1,75E-05 &
  \cellcolor[HTML]{FFEB9C}{\color[HTML]{9C5700} 0.091789} &
  0.005228 &
  0.001614 &
  -0.00051 &
  0.012649 &
  -0.00077 &
  -0.00193 &
  -0.00337 &
  -0.00091 &
  \cellcolor[HTML]{FFC7CE}{\color[HTML]{9C0006} -0.00433} &
  -0.00028 &
  0.000281 &
  -0.00132 &
  0.001982 &
  -0.00028 &
  0.00093 &
  \cellcolor[HTML]{C6EFCE}{\color[HTML]{006100} 0.209298} &
  0.004 \\ \hline
\textbf{IC\_209} &
  -0.00172 &
  0.001333 &
  \cellcolor[HTML]{FFEB9C}{\color[HTML]{9C5700} 0.09286} &
  0.005702 &
  0.01014 &
  0.001281 &
  -0.00239 &
  -0.00354 &
  0.004263 &
  \cellcolor[HTML]{FFC7CE}{\color[HTML]{9C0006} -0.00535} &
  0.000404 &
  -0.00444 &
  -0.00265 &
  0.002825 &
  0.002298 &
  0.007965 &
  -0.00337 &
  0.006035 &
  0.001807 &
  \cellcolor[HTML]{C6EFCE}{\color[HTML]{006100} 0.248175} \\ \hline
\end{tabular}%
}
\caption{Aggregated correlation matrix for images from original camera IC\_192. On the columns the comparison devices on the rows the spoof devices. In green are marked the highest value for comparison in each row, in yellow the second largest value and in red the lowest}
\label{tab:comparisonic192}
\end{table}

The residue \ac{PNU} in these image became more evident analyzing the correlation with the filtered only images (see Table \ref{tab:filteredimages}). In this case, the highest correlation value mean is always associated with the original camera and is almost ten times higher than the rest of the values. The second highest value is randomly distributed over the row.

\begin{table}[htpb]
\centering
\resizebox{\textwidth}{!}{%
\begin{tabular}{|c|l|l|l|l|l|l|l|l|l|l|l|l|l|l|l|l|l|l|l|l|}
\hline
\textbf{} &
  \multicolumn{1}{c|}{\textbf{IC\_190}} &
  \multicolumn{1}{c|}{\textbf{IC\_191}} &
  \multicolumn{1}{c|}{\textbf{IC\_192}} &
  \multicolumn{1}{c|}{\textbf{IC\_193}} &
  \multicolumn{1}{c|}{\textbf{IC\_194}} &
  \multicolumn{1}{c|}{\textbf{IC\_195}} &
  \multicolumn{1}{c|}{\textbf{IC\_196}} &
  \multicolumn{1}{c|}{\textbf{IC\_197}} &
  \multicolumn{1}{c|}{\textbf{IC\_198}} &
  \multicolumn{1}{c|}{\textbf{IC\_199}} &
  \multicolumn{1}{c|}{\textbf{IC\_200}} &
  \multicolumn{1}{c|}{\textbf{IC\_201}} &
  \multicolumn{1}{c|}{\textbf{IC\_202}} &
  \multicolumn{1}{c|}{\textbf{IC\_203}} &
  \multicolumn{1}{c|}{\textbf{IC\_204}} &
  \multicolumn{1}{c|}{\textbf{IC\_205}} &
  \multicolumn{1}{c|}{\textbf{IC\_206}} &
  \multicolumn{1}{c|}{\textbf{IC\_207}} &
  \multicolumn{1}{c|}{\textbf{IC\_208}} &
  \textbf{IC\_209} \\ \hline
\textbf{IC\_190} &
  \cellcolor[HTML]{C6EFCE}{\color[HTML]{006100} 0.077842} &
  -0.00249 &
  -0.00054 &
  -0.00056 &
  \cellcolor[HTML]{FFC7CE}{\color[HTML]{9C0006} -0.00816} &
  -0.00384 &
  \cellcolor[HTML]{FFEB9C}{\color[HTML]{9C5700} 0.001737} &
  -0.00198 &
  -0.00347 &
  -0.0027 &
  -0.00577 &
  -0.00046 &
  -0.00168 &
  -0.002 &
  -0.00121 &
  -0.00374 &
  0.001614 &
  -0.0007 &
  0.00114 &
  -0.00247 \\ \hline
\textbf{IC\_191} &
  \cellcolor[HTML]{FFC7CE}{\color[HTML]{9C0006} -0.00344} &
  \cellcolor[HTML]{C6EFCE}{\color[HTML]{006100} 0.08886} &
  -0.00182 &
  -0.00153 &
  -0.0033 &
  \cellcolor[HTML]{FFEB9C}{\color[HTML]{9C5700} 0.00286} &
  -0.00023 &
  -0.00158 &
  -0.00035 &
  -0.00114 &
  0.000772 &
  0.000491 &
  0.000105 &
  -0.00075 &
  7,02E-05 &
  0.002526 &
  -0.00082 &
  -0.002 &
  -0.00093 &
  -0.00165 \\ \hline
\textbf{IC\_192} &
  -0.00189 &
  -0.00142 &
  \cellcolor[HTML]{C6EFCE}{\color[HTML]{006100} 0.099649} &
  0.002632 &
  -0.00035 &
  -0.00167 &
  -0.00216 &
  -0.00218 &
  -0.0036 &
  -0.00544 &
  -0.00228 &
  \cellcolor[HTML]{FFC7CE}{\color[HTML]{9C0006} -0.00575} &
  -0.00535 &
  0.001877 &
  0.000737 &
  0.001281 &
  -0.00237 &
  \cellcolor[HTML]{FFEB9C}{\color[HTML]{9C5700} 0.003088} &
  -0.00144 &
  0.001561 \\ \hline
\textbf{IC\_193} &
  -0.00102 &
  -0.00135 &
  \cellcolor[HTML]{FFEB9C}{\color[HTML]{9C5700} 0.004316} &
  \cellcolor[HTML]{C6EFCE}{\color[HTML]{006100} 0.092246} &
  -0.00514 &
  -0.00118 &
  0.001368 &
  -0.00314 &
  -0.00388 &
  -0.00274 &
  -0.00189 &
  \cellcolor[HTML]{FFC7CE}{\color[HTML]{9C0006} -0.00696} &
  -0.00418 &
  -0.00061 &
  -0.00265 &
  0.002737 &
  0.001807 &
  0.002947 &
  -0.00012 &
  0.00193 \\ \hline
\textbf{IC\_194} &
  -0.00139 &
  -0.0023 &
  -0.00081 &
  \cellcolor[HTML]{FFC7CE}{\color[HTML]{9C0006} -0.00305} &
  \cellcolor[HTML]{C6EFCE}{\color[HTML]{006100} 0.070982} &
  -0.00086 &
  -0.00182 &
  -0.00081 &
  -0.00107 &
  -0.00014 &
  -0.00107 &
  -0.00168 &
  -0.00175 &
  -0.00133 &
  -0.00251 &
  -0.00167 &
  -0.00168 &
  \cellcolor[HTML]{FFEB9C}{\color[HTML]{9C5700} 0.000193} &
  -0.00081 &
  -0.00193 \\ \hline
\textbf{IC\_195} &
  -0.00282 &
  \cellcolor[HTML]{FFEB9C}{\color[HTML]{9C5700} 0.002579} &
  -0.00044 &
  -0.00037 &
  -0.00207 &
  \cellcolor[HTML]{C6EFCE}{\color[HTML]{006100} 0.068684} &
  -0.00011 &
  -0.00268 &
  -0.00082 &
  -0.00142 &
  -0.0013 &
  \cellcolor[HTML]{FFC7CE}{\color[HTML]{9C0006} -0.00328} &
  -0.00054 &
  -0.0024 &
  -0.00174 &
  0.00186 &
  -0.00042 &
  -0.00219 &
  -0.00074 &
  -0.00026 \\ \hline
\textbf{IC\_196} &
  -7E-05 &
  -8,8E-05 &
  -0.00132 &
  0.000175 &
  -0.00095 &
  -0.00084 &
  \cellcolor[HTML]{C6EFCE}{\color[HTML]{006100} 0.082825} &
  0.000351 &
  -0.00039 &
  -7E-05 &
  -0.00054 &
  -0.00116 &
  \cellcolor[HTML]{FFEB9C}{\color[HTML]{9C5700} 0.000895} &
  -0.00202 &
  \cellcolor[HTML]{FFC7CE}{\color[HTML]{9C0006} -0.00267} &
  -0.00032 &
  -0.00047 &
  1,75E-05 &
  0.000281 &
  -0.00082 \\ \hline
\textbf{IC\_197} &
  -0.00323 &
  -0.00163 &
  -0.00475 &
  -0.00518 &
  \cellcolor[HTML]{FFEB9C}{\color[HTML]{9C5700} 0.005404} &
  -0.00365 &
  0.001947 &
  \cellcolor[HTML]{C6EFCE}{\color[HTML]{006100} 0.105386} &
  0.001456 &
  0.004298 &
  0.003632 &
  0.004807 &
  0.000772 &
  -0.00346 &
  0.001105 &
  -0.00312 &
  \cellcolor[HTML]{FFC7CE}{\color[HTML]{9C0006} -0.00596} &
  -0.00088 &
  -0.00098 &
  -0.00502 \\ \hline
\textbf{IC\_198} &
  -0.00279 &
  -0.00077 &
  -0.00351 &
  \cellcolor[HTML]{FFC7CE}{\color[HTML]{9C0006} -0.00389} &
  \cellcolor[HTML]{FFEB9C}{\color[HTML]{9C5700} 0.005561} &
  -0.00135 &
  0.000421 &
  0.000965 &
  \cellcolor[HTML]{C6EFCE}{\color[HTML]{006100} 0.07707} &
  0.001982 &
  0.000789 &
  0.001298 &
  0.000228 &
  -0.00198 &
  -0.00125 &
  -0.00054 &
  -0.00335 &
  -0.00284 &
  -0.00112 &
  0.000491 \\ \hline
\textbf{IC\_199} &
  -0.00223 &
  -0.00058 &
  \cellcolor[HTML]{FFC7CE}{\color[HTML]{9C0006} -0.00344} &
  -0.00074 &
  \cellcolor[HTML]{FFEB9C}{\color[HTML]{9C5700} 0.002912} &
  -0.00181 &
  -0.00019 &
  0.00086 &
  0.000351 &
  \cellcolor[HTML]{C6EFCE}{\color[HTML]{006100} 0.091737} &
  -0.00233 &
  -0.00153 &
  -0.00102 &
  -0.00105 &
  -0.00107 &
  -0.00156 &
  -0.00088 &
  -0.00228 &
  -0.00095 &
  -0.00182 \\ \hline
\textbf{IC\_200} &
  \cellcolor[HTML]{FFC7CE}{\color[HTML]{9C0006} -0.00375} &
  0.001456 &
  -0.00247 &
  -0.00137 &
  \cellcolor[HTML]{FFEB9C}{\color[HTML]{9C5700} 0.007333} &
  -0.00249 &
  -0.00214 &
  0.002632 &
  0.002298 &
  -0.00018 &
  \cellcolor[HTML]{C6EFCE}{\color[HTML]{006100} 0.093596} &
  -0.00158 &
  -0.0014 &
  0.000263 &
  -0.00056 &
  0.000509 &
  -0.00182 &
  -0.00035 &
  -0.00098 &
  -0.00214 \\ \hline
\textbf{IC\_201} &
  -0.00077 &
  5,26E-05 &
  -0.00095 &
  -0.00121 &
  -0.00261 &
  -0.00347 &
  -0.00079 &
  -0.00104 &
  -8,8E-05 &
  -0.00163 &
  -0.0024 &
  \cellcolor[HTML]{C6EFCE}{\color[HTML]{006100} 0.084228} &
  0.000491 &
  \cellcolor[HTML]{FFEB9C}{\color[HTML]{9C5700} 0.002474} &
  -0.00144 &
  -0.00172 &
  -0.00163 &
  \cellcolor[HTML]{FFC7CE}{\color[HTML]{9C0006} -0.00672} &
  -0.00186 &
  -0.00119 \\ \hline
\textbf{IC\_202} &
  -0.00146 &
  0.000175 &
  -0.00295 &
  -0.00261 &
  0.002877 &
  -0.00137 &
  \cellcolor[HTML]{FFEB9C}{\color[HTML]{9C5700} 0.003754} &
  0.000386 &
  -0.00067 &
  0.001298 &
  0.000333 &
  0.002807 &
  \cellcolor[HTML]{C6EFCE}{\color[HTML]{006100} 0.085351} &
  -0.00267 &
  -0.00196 &
  -0.00028 &
  -0.00279 &
  \cellcolor[HTML]{FFC7CE}{\color[HTML]{9C0006} -0.00458} &
  0.002825 &
  -0.00177 \\ \hline
\textbf{IC\_203} &
  -0.00161 &
  -0.00065 &
  \cellcolor[HTML]{FFEB9C}{\color[HTML]{9C5700} 0.001} &
  -0.00096 &
  -0.00058 &
  -0.00365 &
  \cellcolor[HTML]{FFC7CE}{\color[HTML]{9C0006} -0.00414} &
  -0.00195 &
  -0.00151 &
  -0.00144 &
  7,02E-05 &
  0.000491 &
  -0.00375 &
  \cellcolor[HTML]{C6EFCE}{\color[HTML]{006100} 0.086158} &
  -0.0003 &
  -0.00119 &
  -0.00139 &
  0.000649 &
  -0.00407 &
  -0.00112 \\ \hline
\textbf{IC\_204} &
  -0.00139 &
  -0.00088 &
  0.000702 &
  -0.0013 &
  \cellcolor[HTML]{FFC7CE}{\color[HTML]{9C0006} -0.00267} &
  -0.00198 &
  -0.00163 &
  0.000456 &
  -0.0006 &
  0.000175 &
  -0.00163 &
  -0.00012 &
  0.000263 &
  -0.0003 &
  \cellcolor[HTML]{C6EFCE}{\color[HTML]{006100} 0.104912} &
  -0.00063 &
  -0.0013 &
  \cellcolor[HTML]{FFEB9C}{\color[HTML]{9C5700} 0.001667} &
  -0.00082 &
  5,26E-05 \\ \hline
\textbf{IC\_205} &
  -0.00239 &
  0.001544 &
  0.001158 &
  0.001579 &
  0.000439 &
  0.000737 &
  -0.00037 &
  -0.00172 &
  0.000509 &
  -0.00086 &
  -0.00058 &
  \cellcolor[HTML]{FFC7CE}{\color[HTML]{9C0006} -0.00572} &
  -0.00042 &
  -0.00072 &
  -0.00114 &
  \cellcolor[HTML]{C6EFCE}{\color[HTML]{006100} 0.091105} &
  -0.00025 &
  -0.0003 &
  -0.00091 &
  \cellcolor[HTML]{FFEB9C}{\color[HTML]{9C5700} 0.002263} \\ \hline
\textbf{IC\_206} &
  -0.00109 &
  -0.00147 &
  -1,8E-05 &
  \cellcolor[HTML]{FFEB9C}{\color[HTML]{9C5700} 0.001035} &
  \cellcolor[HTML]{FFC7CE}{\color[HTML]{9C0006} -0.00718} &
  -0.00096 &
  8,77E-05 &
  -0.00381 &
  -0.00365 &
  -0.00291 &
  -0.00046 &
  -0.00474 &
  -0.00349 &
  -0.00111 &
  -0.00354 &
  -0.00056 &
  \cellcolor[HTML]{C6EFCE}{\color[HTML]{006100} 0.082053} &
  0.000228 &
  -0.00133 &
  -0.00067 \\ \hline
\textbf{IC\_207} &
  -0.00132 &
  -0.00105 &
  0.000509 &
  0.000965 &
  0.001158 &
  -0.00186 &
  0.000316 &
  -0.00109 &
  -0.00139 &
  -0.00279 &
  -0.00084 &
  \cellcolor[HTML]{FFC7CE}{\color[HTML]{9C0006} -0.00896} &
  -0.00311 &
  -0.00098 &
  \cellcolor[HTML]{FFEB9C}{\color[HTML]{9C5700} 0.003947} &
  -0.00104 &
  0.000474 &
  \cellcolor[HTML]{C6EFCE}{\color[HTML]{006100} 0.097386} &
  -0.00193 &
  0.000912 \\ \hline
\textbf{IC\_208} &
  -0.00032 &
  -0.00023 &
  -0.00102 &
  0.000702 &
  -0.00019 &
  -0.00149 &
  \cellcolor[HTML]{FFEB9C}{\color[HTML]{9C5700} 0.002439} &
  -0.00035 &
  -0.00049 &
  -0.00011 &
  0.000702 &
  -0.00242 &
  0.001737 &
  \cellcolor[HTML]{FFC7CE}{\color[HTML]{9C0006} -0.00282} &
  -0.00154 &
  -0.00088 &
  -0.00068 &
  -0.00158 &
  \cellcolor[HTML]{C6EFCE}{\color[HTML]{006100} 0.098351} &
  -0.00116 \\ \hline
\textbf{IC\_209} &
  -0.00239 &
  -0.00021 &
  0.00086 &
  0.001509 &
  \cellcolor[HTML]{FFEB9C}{\color[HTML]{9C5700} 0.004474} &
  -0.00116 &
  8,77E-05 &
  -0.00265 &
  0.001053 &
  0.000246 &
  -0.00156 &
  \cellcolor[HTML]{FFC7CE}{\color[HTML]{9C0006} -0.00465} &
  0.000368 &
  -0.00125 &
  -0.00088 &
  0.001702 &
  -0.00063 &
  0.001246 &
  -0.00037 &
  \cellcolor[HTML]{C6EFCE}{\color[HTML]{006100} 0.076439} \\ \hline
\end{tabular}%
}
\caption{Aggregated correlation matrix for filtered only images. On the columns the comparison devices on the rows the original devices. In green are marked the highest value for comparison in each row, in yellow the second largest value and in red the lowest}
\label{tab:filteredimages}
\end{table}

\section{Conclusions and Future Works}
\label{sec:conclusions}
In this paper, we have analyzed some of the existing attack procedures in SCI and developed a different approach for the search for image counterfeits through fingerprints
that achieve a good trade-off between the requirements described in Section 4. Indeed, the proposed method required physical access to the source camera device and the target camera is must be the same model as the original one. 
Therefore, though the experimental results obtained still have some limitations due to certain circumstances, they continued to confirm the robustness of the \ac{PRNU}.

Future works will be dedicated to investigate other smarter attacks from hacker's side on the \ac{PRNU} through the use of smartphone technologies. In fact, the latest generation smartphones are, extremely, powerful and versatile. Developed according to new technologies, most of them are equipped with multiple cameras and allow you to take excellent images. All this thanks to a new photo generation process that has more details and noise reduction, through its reconstruction that takes the best pixels from the other shots. In this new scenario, it will be interesting to verify the existence and robustness of the PRNU.

%
%
%
\bibliographystyle{splncs04}
%
\bibliography{bibliography}
\end{document}